\documentclass[review,authoryear,12pt]{elsarticle}

 \usepackage{graphicx}

\usepackage{amssymb}

\journal{Icarus}

\begin{document}

\begin{frontmatter}

\title{Tectonic Activity on Pluto After the Charon-Forming Impact}

\author[label1]{Amy C. Barr}
\author[label2]{Geoffrey C. Collins}

 \address[label1]{Brown University, Dept. of Geological Sciences, Providence, RI 02912, USA}
 \address[label2]{Department of Physics and Astronomy, Wheaton College, Norton, MA 02776, USA}

\begin{abstract}
The Pluto-Charon system, likely formed from an impact, has reached the endpoint of its tidal evolution. During its evolution into the dual-synchronous state, the equilibrium tidal figures of Pluto and Charon would have also evolved as angular momentum was transferred from Pluto's spin to Charon's orbit.  The rate of tidal evolution is controlled by Pluto's interior physical and thermal state.  We examine three interior models for Pluto: an undifferentiated rock/ice mixture, differentiated with ice above rock, and differentiated with an ocean.  For the undifferentiated case without an ocean, the Pluto-Charon binary does not evolve to its current state unless its internal temperature $T_i>200$ K, which would likely lead to strong tidal heating, melting, and differentiation.  Without an ocean, Pluto's interior temperature must be higher than 240 K for Charon to evolve on a time scale less than the age of the solar system.  Further tidal heating would likely create an ocean.  If \emph{New Horizons} finds evidence of ancient tidally-driven tectonic activity on either body, the most likely explanation is that Pluto had an internal ocean during Charon's orbital evolution.
\end{abstract}


\end{frontmatter}


\clearpage
\section{Introduction}  \label{sec:intro_section}

As the \emph{New Horizons} spacecraft travels toward the Pluto/Charon system, our knowledge about the system's physical properties \citep{Olkin2003, Gulbis2006, Tholen2008}, mode of formation \citep{Canup2005,Canup2011}, and surface composition (e.g., \citealt{Olkin2003, BrownCalvin2000, BuieGrundy2000, Cook2007}) continues to grow.  The general properties of the Pluto and Charon system, including its dynamical state and the Pluto-to-Charon mass ratio, suggest that Charon may have formed due to a collision between two like-sized precursor objects \citep{Canup2005,Canup2011}.  Careful examination of the system in preparation for the \emph{New Horizons} mission has led to the discovery of four additional small moons, Nix, Hydra \citep{Weaver2006}, Styx, and Kerberos \citep{Showalter2011,Showalter2012}.  The moons are in mean motion resonances with Charon \citep{Weaver2006, Stern2006, Buie2006, Showalter2013}, which could imply that they formed by accretion of debris after the Charon-forming impact or were captured \citep{Ward2006, KenyonBromley2013}.

The Pluto/Charon system is unique among major bodies in the solar system because it has reached the endpoint of its dynamical evolution: the so-called ``dual synchronous'' state in which Charon's orbital period, spin period, and Pluto's rotation period are equal; Charon currently orbits Pluto at $a_{c} \sim 16.4 R_P$, where $R_P=1147$ km is Pluto's radius \citep{Tholen2008}.  As the orbit evolved to its present state and Pluto's spin rate changed to match Charon's orbital period, changes in the equilibrium tidal and rotational figures of the two bodies may have left their mark as systems of tectonic features on the surfaces of each body.  Orbit and spin evolution of the Pluto/Charon system is driven by the raising and lowering of tidal bulges on each body.  The tidal bulges exert torques which change the semi-major axes of the orbits and the spin rates. The mechanical energy associated with the periodic raising and lowering of the bulges is dissipated as heat in the bodies' interiors.  Changes in bulge height and energy dissipation are thought to drive endogenic resurfacing and tectonic activity on many of the icy satellites of the outer solar system (see, e.g., \citealt{SchubertSatsBook, PealeSatsBook,EuropaJupBook} for discussion).  The magnitude of  stresses arising from tidal evolution depends on the interior structures of the bodies and the frequency of tidal flexing; the frequencies in turn depend on the Pluto/Charon distance and the spin periods.  The main source of stress accumulation in the Pluto/Charon system would be figure changes associated with transfer of momentum from Pluto's spin to Charon's orbit.  If the stresses exerted on the surfaces of the bodies exceed the nominal yield stress of ice, we consider that tectonic activity may have occurred.  The presence or absence of tectonic features, along with their distribution and orientations, may provide clues about the early evolution of the system \citep{CollinsPappalardoLPSC2000}.  Thus, images of the surfaces of Pluto and Charon from \emph{New Horizons} could yield clues about the post-impact interior state of Pluto and the initial orbital distance of Charon.  

In successful hydrodynamical simulations of the Charon-forming impact, Charon is launched into an eccentric orbit around a rapidly spinning Pluto (cf. \citealt{DobroPlutoBook}); its initial orbital semi-major axis after the impact, $a_o$, ranges from $\sim 3.7R_P$ to $21 R_P$ \citep{RobinPers}.  Over a time scale $\tau_{synch} \sim 35 (Q_c/100)$ years \citep{DobroPlutoBook}, where $Q_c\sim 100$ is a nominal estimate of Charon's tidal quality factor, Charon evolves to a synchronous state in which its spin period and orbital period are equal (similar to the Earth-Moon system).  The final migration from the synchronous state to the dual-synchronous state takes much longer, $\Delta t_{evol} \approx 200 (Q_p/100)(10^{-3}/k_{2,p})$Myr, where $Q_p$ is Pluto's tidal quality factor, and $k_{2,p}$ is the degree-2 Love number that describes how Pluto's gravitational potential changes in response to the tides raised on Pluto by Charon.

In this work, we calculate the migration timescale and stresses generated in Pluto's lithosphere due to the orbital evolution of the system from its initial post-impact state to its present dual-synchronous state.  We determine the Love numbers of Pluto as a function of its interior structure, temperature, and the time scale of deformation.  The Love numbers are used to estimate its $Q$ immediately post-impact, and to determine the magnitude of tidal deformation in a rapidly spinning Pluto. The Love numbers are also used to constrain the magnitude of deformation and stresses built up in Pluto's lithosphere.  The $Q$ values based on realistic interior structures for Pluto are used to constrain the orbital evolution time scale of the system and the rate of deformation associated with decreasing Pluto's tidal bulge as Charon recedes. Assuming nominal parameters describing the brittle and ductile behavior of water ice (e.g., Table \ref{table:lovenumber_params}), we determine the conditions under which the induced stresses and deformation rates can fracture Pluto's surface, and determine how that likelihood varies as a function of Pluto's thermal state post-impact.


\section{Background}  \label{sec:background}

The discovery of new satellites in the Pluto system has provided new constraints on the masses of Pluto and Charon.  Table \ref{table:satellite_params} summarizes the present-day knowledge of the physical properties of the system based on a four-body orbital solution from \citet{Tholen2008}, which also incorporates prior information about Pluto's radius from stellar occultation data (e.g., \citealt{Olkin2003, Gulbis2006}).  The current estimate for the mean density of Pluto is $\bar{\rho_p}=2060$ kg m$^{-3}$ \citep{Tholen2008, Canup2011}.  The spin/orbital period of Charon at present is 6.38720 days \citep{Tholen2008}.  In the dual-synchronous state, the spin frequency of Pluto ($\omega_p$), Charon ($\omega_c$), and orbital frequency of Charon ($n_f$) are the same, so that $\omega_p=\omega_c=n_f=1.139 \times 10^{-5}$ s$^{-1}$.  Charon orbits Pluto at a distance of $a_f=16.4R_p$.

It is commonly held that Charon formed in an impact between two like-sized precursor ice/rock bodies \citep{Canup2005,Canup2011}.  Simulations of the impact which produce the proper mass and mean density for Charon show that the object striking proto-Pluto must be undifferentiated at the time of impact to create Charon's mixed ice/rock composition \citep{Canup2005}.  No such constraints exist for the differentiation state of Pluto; a collision between undifferentiated objects is consistent with the present-day masses and densities of each object.  The energy rise associated with the impact is modest, with the interior of Pluto being heated by $\Delta T \sim O(10-100)$ K, and Charon's interior being heated by $\Delta T \sim 30$ K \citep{Canup2005}.  It appears likely that the interiors of either object could be heated to within a few tens of degrees Kelvin of their melting points by long-lived radiogenic heating before the impact, particularly if their interiors contain ammonia, which would decrease the melting points of their icy components \citep{Hussmann2006,McKinnon_KBO_Book}.

The evolution of the binary system to the dual-synchronous state is controlled by the tides raised on Pluto due to Charon's orbit \citep{DobroPlutoBook}.  If Charon's $a_o< 16.4R_p$, angular momentum is transferred from Pluto's spin to Charon's orbit, causing Charon to evolve outward as Pluto de-spins.  If Charon's $a_o > 16.4 R_p$, angular momentum is transferred from Charon's orbit to Pluto's spin, and Charon moves closer to Pluto until it achieves the dual-synchronous state.  Tides within Pluto are raised and lowered with a frequency $\omega_{tidal}=\omega_{p}-n$, where $\omega_p$ and $n$ are the spin frequency of Pluto and Charon's mean motion, respectively, both of which change during orbital evolution.  

Immediately following the Charon-forming impact, we assume that Charon is launched into a highly eccentric, co-planar orbit around a rapidly spinning Pluto (cf. \citealt{DobroPlutoBook}); this eccentricity is damped rapidly.  Assuming a constant tidal quality factor for Pluto, the change in Charon's mean motion due to the torque between Pluto and Charon is related to their physical properties as \citep{DobroPlutoBook},
\begin{equation}
\frac{dn}{dt} = - \frac{9}{2} \frac{k_{2,p}}{Q} \frac{m_c}{m_p} \frac{R_p^5 n^{16/3}}{[G (m_p + m_c)]^{5/3}}, \label{eq:dndt_dobro}
\end{equation}
where $m_p$ is Pluto's mass, $m_c$ is Charon's mass, $k_{2,p}$ is Pluto's degree-2 Love number describing the change in Pluto's gravitational potential due to its distortion, and $Q$ is Pluto's tidal quality factor.  Assuming a constant Q as a function of time, equation (\ref{eq:dndt_dobro}) can be integrated to give the time scale of orbital evolution for the system (cf. \citealt{DobroPlutoBook}),
\begin{equation}
\tau_{evol} = \frac{2}{39} \frac{Q}{k_{2,p}} \frac{m_p}{m_c} \frac{[G(m_p+m_c)]^{5/3}}{R_p^5} (n_f^{-13/3} - n_o^{-13/3}) , \label{eq:approximate_tevol}
\end{equation}
where $n_o=(G m_p/a_o^3)^{1/2}$ is the mean motion of Charon immediately post-impact.   For the minimum $a_o/R_p \sim 3.7$ (which gives $n_o \sim 10^{-4}$ s$^{-1}$), nominal estimates for $Q$ and $k_{2,p}$, and $n_f=1.13 \times 10^{-5}$ s$^{-1}$, equation (\ref{eq:approximate_tevol}) gives $\tau_{evol} \sim  200$ Myr.  As discussed in section \ref{sec:Orbevol}, the value of $Q$ is dependent on the frequency of tidal deformation and the interior state of Pluto.  Therefore, an overall $\tau_{evol}$ based on a constant $Q$ should be regarded only as an initial estimate that does not use the full information available.


A crude estimate of the tidal distortion of Pluto during its early period of rapid rotation can shed light on the stresses exerted in its lithosphere during this early phase of orbital evolution.  The magnitude of the strain, $\varepsilon$, on Pluto's lithosphere due to the tidal deformation is also related to $k_{2,p}$ and the system properties \citep{Love,DobroPlutoBook},
\begin{equation}
\varepsilon = \frac{\delta r}{R_p} = \frac{5}{3} k_{2,p} \frac{m_c}{m_p} \bigg(\frac{R_p}{a_o}\bigg)^3 \sim 3 \% \bigg( \frac{k_{2,p}}{0.001}\bigg),  \label{eq:approximate_stresses}
\end{equation}
for $a_o \sim 3.7 R_p$ and a nominal $k_{2,p} \sim O(10^{-3})$ \citep{DobroPlutoBook}.  This implies stresses of order $\sigma \sim 2 \mu \varepsilon \sim 100$ MPa, where $\mu = 3.6 \times 10^9$ Pa is the shear modulus for water ice.  These stresses are significantly higher than the $\sim 0.1$ MPa yield stress of polycrystalline water ice inferred from terrestrial ice sheets \citep{Kehle1964}, and higher than the yield stress of the lithosphere of Europa inferred from the models of the formation of its tectonic features \citep{Hoppa99,Stempel2005,HurfordCycloids2007}.  It is also higher than the $\sim 2$ MPa tensile strength of polycrystalline water ice at temperatures found in the outer solar system \citep{Litwin2013}.  This suggests that stresses from figure change may be sufficient to fracture Pluto's surface and create tectonic features \citep{CollinsPappalardoLPSC2000} that could be observable by \emph{New Horizons}.


\section{Methods}
Here, we estimate the time scale of orbital evolution from the synchronous to dual-synchronous state as a function of the interior viscosity of Pluto for three different interior states: a differentiated Pluto with an ocean, a differentiated Pluto without an ocean, and an undifferentiated Pluto (see Figure \ref{fig:interiors}).  We derive a time estimate by integrating the constant-$Q$ approximation in equation (\ref{eq:approximate_tevol}) for small steps in the Pluto-Charon distance, varying $Q$ as a function of distance.  We calculate Love numbers for Pluto assuming a radially symmetric structure and vary the thermal state by using different values for the viscosity of ice in the interior.  Finally, we calculate the magnitude of surface stresses available to drive tectonic activity by assuming deformation of a thin elastic shell \citep{Melosh77} due to changes in the equilibrium figures of Pluto and Charon as orbital evolution proceeds.

\subsection{Interior Models} \label{sec:interiors}
Figure \ref{fig:interiors} illustrates the three simple models we consider for the interior structure of Pluto immediately after the Charon-forming impact: a homogeneous (uniform-density) interior, a fully differentiated and solid interior, and a fully differentiated interior with an ocean.  Table \ref{table:satellite_params} summarizes the physical properties of Pluto.  

We assume that Pluto is composed primarily of water ice with a density $\rho_i=1000$ kg m$^{-3}$, slightly higher than the density of cold, pure water ice \citep{MuellerMcKinnon} to account for the possible inclusion of hydrocarbons and higher-density N-bearing compounds \citep{McKinnonPlutoBook,McKinnon_KBO_Book}.  We assume that its rocky component has a density $\rho_r=3000$ kg m$^{-3}$, mid-way between the grain density of CI chondrite ($\rho_r=2800$ kg m$^{-3})$ and Prinn/Fegley rock $(\rho_r=3300$ kg m$^{-3})$, which are two models for the composition of the rocky components of large icy moons \citep{MuellerMcKinnon}.  With our choices of rock and ice densities, Pluto has a volume fraction of rock $\phi=(\bar{\rho}-\rho_i)/(\rho_r - \rho_i)=0.53$, consistent with more detailed structural models of McKinnon et al., (2008).  

Not much is known about the interior state of Pluto before the Charon-forming impact.  Much of the prior work on Pluto's interior assumes that it differentiated during its formation and early history \citep{McKinnonPlutoBook, McKinnon_KBO_Book, RobuchonNimmo}. However, simulations of the Charon-forming impact show that the proper mass, composition, and angular momentum for the system is most readily obtained when Pluto and the Charon-forming impactor are only partially differentiated \citep{Canup2005,Canup2011}.

Here, we consider both cases as a possibility, but we favor differentiated Plutos. The gravitational binding energy associated with assembling Pluto, $3/5 (GM_P/R_P) \sim 4.5 \times 10^5$ J/kg (where $M_P$ is Pluto's mass) is larger than the latent heat of water ice, $L \sim 3 \times 10^5$ J/kg, suggesting that Pluto's ice could melt during its accretion if 100\% of its accretional energy were retained as heat.  This may not be the case, especially if Pluto formed slowly, and from low-velocity impacts (see e.g., \citealt{AhrensOKeefe1977}).  If Pluto formed very early in solar system history, before the dissipation of the solar nebula ($\sim 10$ Myr), comparable to the lifetime of disks around other stars \citep{Haisch2001, Thi}, it may have been heated by the short-lived radioisotopes (SLRI) $^{26}$Al and $^{60}$Fe, which have half-lives less than $\sim 1$ Myr and are present in greater abundances than U, Th, and K \citep{TachibanaHuss2003, Lodders2003, Mostefaoui2004, Bizzarro2005, Thrane2006,  Wadhwa2007}.  SLRI are particularly efficient at driving differentiation because they release heat on a time scale much shorter than the time scale for diffusive or convective heat transport in Pluto's interior.  Removing a heat pulse from short-lived radioisotope decay across the outer $\delta \sim 500$ km of Pluto's interior would require $\tau_{diff} \sim \delta^2/\kappa \sim 7$  Gyr, where $\kappa \sim 10^{-6}$ m$^2$/s is the thermal diffusivity of water ice.  Solid-state convection could provide more efficient heat transport, but the time scale for the first convective overturn in Pluto's deep interior \citep{Zaranek2004},
\begin{equation}
t_o = \frac{500}{\kappa} \bigg(\frac{\rho \alpha_T \Delta T_i g}{\eta_o \kappa}\bigg)^{-2/3},
\end{equation}
where $\alpha_T \sim 10^{-4}$ K$^{-1}$ is the coefficient of thermal expansion of ice, $\Delta T_i \sim 20$ K is the approximate magnitude of temperature variations driving convection in Pluto's core,  $g$ is the local acceleration of gravity on Pluto, and $\eta_o$ is the viscosity of ice in Pluto's deep interior \citep{cr}.  A lower limit on the time scale for the onset of convection $t_o \sim 2$ Myr can be obtained using the melting point viscosity of ice I, $\eta_o=10^{14}$ Pa s, and Pluto's surface gravity $g=0.65$ m/s$^2$.  Because $t_o$ is comparable to the half-lives of $^{26}$Al and $^{60}$Fe, it would be difficult for Pluto to remove internal heat from these sources before they could drive internal melting.  Once differentiation had begun inside Pluto, it would have become an energetically self-sustaining process, driving itself to completion.  This is because the gravitational potential energy per unit mass associated with the differentiation of Pluto \citep{FriedsonStevenson, lhbpaper},
\begin{equation}
E_{gr} = \frac{3}{5} \frac{GM}{R_p} \bigg[1-z_1^2 x^5 - z_2^2 (1-x^5) - \frac{5}{2}x^3 z_2 (z_1-z_2)(1-x^2)\bigg]
\end{equation}
where $z_1=\rho_r/\bar{\rho}$, $z_2=\rho_i/\bar{\rho}$, $x=\phi^{1/3}$, and $E_{gr} \sim 5 \times 10^5$ J/kg, is larger than the latent heat of water ice.

We also consider it likely that Pluto had an ocean post-impact, and that the ocean may persist even to the present day.  Assuming an equilibrium between present chondritic heat flow and the convective heat flux, Pluto could be warm enough to harbor an ocean beneath $> 100$ km of ice \citep{Hussmann2006}. Variations in ice shell thickness can lead to stresses that differ by a factor of $\sim 2$ from estimates here.  However, we seek estimates of stress at an order of magnitude level, so we do not explore the effect of shell thickness variations here.  Thicker oceans are possible in the past, when the chondritic heat flux was higher, and if some kind of low-eutectic material (such as ammonia) is mixed in with Pluto's ices \citep{Hussmann2006}.  Of course, as Charon's orbital evolution results in tidal heating inside Pluto, the combination of impact induced heat, radiogenic heat, and tidal heat is likely to drive melting, which creates an ocean and could trigger runaway differentiation if Pluto were undifferentiated before.  Although these simple arguments favor a differentiated Pluto with an ocean, we additionally explore cases for Pluto evolving with no ocean, and with a uniform-density interior, because they are both permitted by the impact simulations \citep{Canup2005, Canup2011}.

\subsection{Love Numbers} \label{sec:Lovenumbers}
As described in Section \ref{sec:background}, the orbital evolution time scale for the system is strongly dependent on the $Q$ of Pluto, which expresses the fraction of orbital energy dissipated during each diurnal tidal cycle.  Previous studies have considered $Q=100$ (e.g., \citealt{DobroPlutoBook}), a value commonly assumed for icy satellites \citep{MurrayDermott}.  Models of the tidal deformation of Europa and Enceladus \citep{RossSchubert89,MooreSchubert,Wahr2006} show that the values of $k_2$ and $Q$ for a mixed ice/rock body depend on the degree of differentiation and whether the object has an ocean decoupling its outer ice shell from its deeper interior.  Thus, the time scale for system evolution and the potential for generating tidally induced tectonics on Pluto depend strongly on its post-impact interior structure.

We estimate $Q$ by determining how the interior of a spherically symmetric viscoelastic Pluto deforms in response to a tidal bulge raised due to Charon on a frequency $\omega_{tidal}$.  We use numerical techniques to calculate $k_2$ using the correspondence principle, an approach has been used to study tidal heating in rocky and icy bodies including Io, Europa, and Enceladus \citep{Segatz1988, RossSchubert89, MooreSchubert, Wahr2006, NimmoEncelShear, encel_sluggish, SatStress}.  Details regarding the particular methods we employ, namely the portion of \emph{SatStress} that calculates Love numbers for viscoelastic planetary bodies, may be found in \citet{SatStress}.  In this approach, the response of a spherically symmetric viscoelastic planet to an applied tidal potential with varying harmonic degree is calculated as a function of the tidal frequency and material properties of the planet.  

For a viscoelastic body, $k_2$ can be considered a complex number (which we denote $k_2^*$), where the real part of $k_2^*$, $Re(k^*_2)$, is related to elastic (recoverable, adiabatic) deformation occurring over a time scale $2 \pi/\omega_{tidal}$, and the imaginary part, $Im(k^*_2)$, is related to the non-recoverable dissipation occurring inside Pluto over the tidal time scale \citep{Segatz1988, SatStress}.  The tidal quality factor is related to $k^*_2$ \citep{Segatz1988},
\begin{equation}
Q=\frac{|k^*_2|}{Im(k_2^*)},
\end{equation}
which allows us to substitute $1/Im(k_2^*)$ for the ratio $Q/k_2$ in equation (\ref{eq:approximate_tevol}).

\subsubsection{Interior States}
\begin{figure}
\includegraphics[type=eps,ext=.eps,read=.eps,scale=0.50]{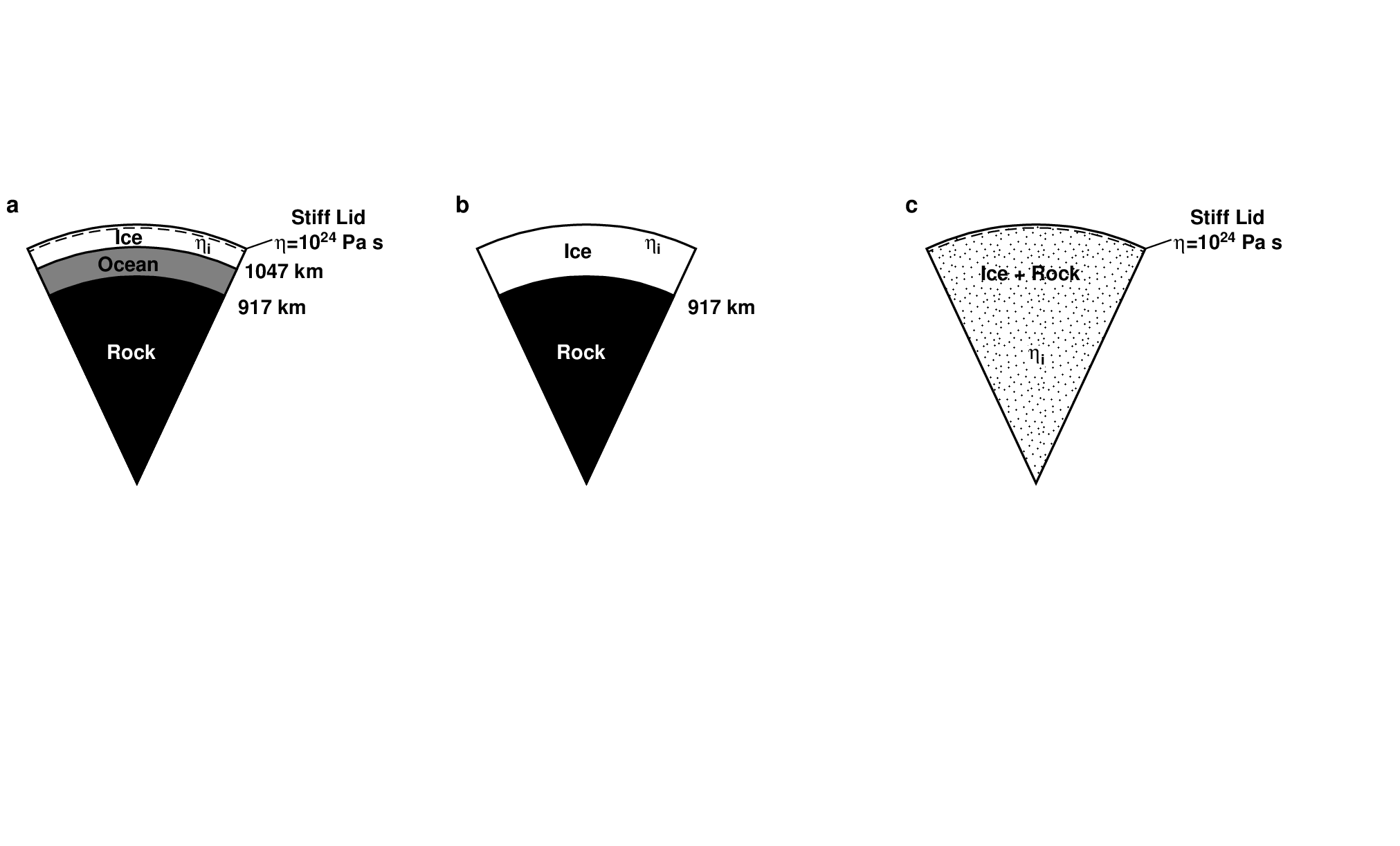}
\caption[interiors]{Interior structure models assumed for Pluto.  (a) Fully differentiated with a rocky core of density $\rho_r=3000$ kg m$^{-3}$, liquid ocean of density $\rho_w=1000$ kg m$^{-3}$, and 100 kilometer thick outer ice shell with density $\rho_i=1000$ kg m$^{-3}$  The ice shell has a 15 kilometer-thick layer of stiff ice (effective viscosity $\eta=10^{24}$ Pa s); the viscosity of the lower portion of the ice shell $\eta_i$ is a key control on the amount of tidal deformation and stress experienced by Pluto during Charon's orbital evolution.  The relative thickness of solid ice compared to liquid is unknown, but the amplitude of the tidal deformation is mainly controlled by the presence of the ocean, and is relatively insensitive to the thickness of the ice shell \citep{Wahr2006}.  (b) Fully differentiated Pluto with no ocean, where the inner rock core is frozen to, and thus rigidly coupled to, the overlying ice shell. The viscosity of the ice shell $\eta_i$ controls the amount of tidal deformation. (c) A uniform-density Pluto.  The model has a stiff lid with effective viscosity $\eta=10^{24}$ Pa s, and the viscosity of the interior $\eta_i$ controls the amount of tidal deformation.  \label{fig:interiors}}
\end{figure}

For a differentiated Pluto with an ocean (Figure \ref{fig:interiors}a), we consider an elastic rocky core and an overlying viscoelastic ice shell.  The values of Young's modulus and Poisson's ratio for rock are summarized in Table \ref{table:lovenumber_params}.  We use an ice shell with a thickness of 100 km \citep{Hussmann2006} and an upper layer of cold rigid ice 15 km thick, to represent the cold lithosphere atop the warm sublayer of a convecting ice shell.  We assume a Newtonian volume diffusion rheology for ice, appropriate for deformation at low stresses \citep{GoldsbyKohlstedt, gsevol}.  The viscosity varies as a function of temperature as \citep{GoldsbyKohlstedt},
\begin{equation}
\eta(T) = \frac{3 R_G T d^2}{14 V_m D_{o,v}} \exp \bigg(\frac{Q^*}{R_G T} \bigg) \label{eq:rheology}
\end{equation}
with gas constant $R_G=8.314$ J/mol-K, molar volume $V_m=1.95 \times 10^{-5}$ m$^3$/mol, diffusion constant $D_{o,v}=9.10 \times 10^{-4}$ m$^2$/s, and activation energy $Q^*=59.4$ kJ/mol.  The viscosity is strongly dependent on ice grain size, $d$.  Given the low stresses associated with flow in the interior of Pluto, and the likely presence of small silicate particles mixed in its ices, the grain size in Pluto's interior is plausibly $d \sim 0.1$ to 1 mm \citep{gsevol}.  This yields a melting point viscosity for water ice of $\eta = 5 \times 10^{14} (d/\textrm{0.3 mm})^2$ Pa s, or $\eta \sim 6 \times 10^{13}$ Pa s for $d=0.1$ mm, close to the lower limit of what could be expected in natural systems \citep{gsevol}.  To explore the full range of possible values of ice viscosity, we vary the viscosity of the lower ice layer from $6 \times 10^{13}$ Pa s to $10^{20}$ Pa s.  We consider $10^{20}$ Pa s a plausible upper limit for the ``bulk'' viscosity of a convective or conductive ice shell on Pluto, corresponding to the viscosity of ice at $T\sim 200$ K.  The cold ice ``lithosphere'' has a constant viscosity of $10^{24}$ Pa s, corresponding to a mean temperature $T\sim 150$ K.  The ocean is represented by a fluid layer with a density $\rho_w=1000$ kg/m$^3$.  

The interior state for a differentiated Pluto without an ocean (Figure \ref{fig:interiors}b) is similar to the differentiated case (see Table \ref{table:lovenumber_params}), but without the fluid layer.  We only consider a single ice layer in this case, to obtain upper limits on the value of $k^*_2$ and a lower limit on $Q$ and thus the timescale for orbital evolution for Charon.  If we added a near-surface of layer of cold ice with $\eta \sim 10^{24}$ Pa s, this would only serve to decrease $k_2^*$.

For a uniform-density Pluto, we assume that the interior consists of a convecting homogeneous ice/rock mixture with a constant viscosity, overlain by a 15 kilometer thick ice ``lithosphere.''  We assume that the mixture of ice and rock in the convecting sublayer has a Young's modulus and Poisson's ratio similar to those for pure water ice.  The viscosity of the sublayer is varied from $6 \times 10^{13}$ Pa s to $10^{20}$ Pa s.  The viscosity of ice with up to $\sim 40$ to $50$\% silicate by volume is only a factor of $\sim 2$ higher than the viscosity of pure water ice \citep{FriedsonStevenson, DurhamParticulates}.

\subsubsection{Tidal Frequency}
Because angular momentum is conserved during the orbital evolution of the system, the present total angular momentum of the system can be used to relate Pluto's initial spin frequency, $\omega _p$, to the initial semi-major axis of Charon's orbit, $a_o$.  The total angular momentum of the present system is the sum of the orbital angular momentum of Charon and the spin angular momentum of each body, $L_{tot}=L_{orbital} + L_{spin,p} + L_{spin,c}$, where $L_{orbital}= n_f a^2 (m_p m_c)/(m_p+m_c)$ and $L_{spin}=\alpha m r^2 \omega$, where $\alpha$ is the moment of inertia coefficient, $\alpha=C/mr^2$, and $\omega=n_f$ because Pluto's spin frequency, Charon's spin frequency, and Charon's orbital frequency are presently equal.  The angular momentum immediately following the impact has a similar form, with $\omega_c = n_o = (Gm_p/a_o^3)^{1/2}$ with $a_o$ as a free parameter.  

Equating the present angular momentum equal to that immediately post-impact, we obtain a relationship between the spin frequency of Pluto post-impact, Charon's initial orbital semi-major axis, and the current dynamical state of the system,
\begin{equation}
\frac{\omega_p}{n_f} =1+ \frac{m_c}{\alpha_p(m_p+m_c)} \bigg[\bigg(\frac{a_f}{r_p}\bigg)^2 - \bigg(\frac{a_o}{r_p}\bigg)^2 \bigg(\frac{n_o}{n_f}\bigg)\bigg]   + \frac{\alpha_c}{\alpha_p} \frac{m_c}{m_p} \bigg(\frac{r_c}{r_p}\bigg)^2 \bigg(1-\frac{n_o}{n_f}\bigg).  \label{eq:initial_pluto_omega}
\end{equation}
The initial tidal frequency $\omega_{tidal} = \omega_p -n_o$.  We calculate $k_2^*$ for Pluto for every value of $a_o$ from 3.7 $r_p$ to 21 $r_p$, in increments of 0.1 $r_p$, for mean interior viscosity values from $\sim 6 \times 10^{13}$ to $10^{20}$ Pa s, in five logarithmically spaced increments for each order of magnitude.  

The moment of inertia coefficient ($\alpha$) for each body is \citep{FriedsonStevenson},
\begin{equation}
\alpha = \frac{2}{5} \frac{1}{\bar{\rho}} \bigg[\rho_i (1-x^5) + \rho_r x^5\bigg],
\end{equation}
where $x$ is the fractional radius of the rock core in each body.  When Pluto is fully differentiated, we assume that Charon is as well; for uniform-density plutos, we assume that Charon is undifferentiated.  

\subsection{Orbital Evolution Due to Pluto Tides} \label{sec:Orbevol}

For each combination of interior structure and $k_2^*$ calculated from $a_o$ and mean interior viscosity, equation (\ref{eq:approximate_tevol}) can be used to determine the time it takes for Charon's orbit to move by 0.1 $r_p$, which is the spacing between successive calculations of $k_2^*$ in section \ref{sec:Lovenumbers}. We then integrate equation (\ref{eq:approximate_tevol}) for all Charon starting distances and Pluto viscosities by summing the time spent at all intermediate stages from $a_o$ to $a_f$.

\subsection{Thermal Evolution of Pluto}
Although the total amount of tidal energy dissipated during the orbital evolution is modest \citep{DobroPlutoBook}, the heat \emph{flux} through the ice shell could be substantial, particularly if the orbital evolution time scale is short.  The average energy dissipation rate within Pluto over the time period of orbital evolution is the sum of the change in Pluto's rotational energy and the system orbital energy, as angular momentum is conserved. If eccentricity is zero, the energy dissipation rate within Pluto is 
\begin{equation}
\frac{dE_p}{dt} =\frac{1}{2}\bigg(\omega_p m_c \sqrt{\frac{G m_p}{a}} + \frac{G m_p m_c}{a^2}\bigg)\frac{da_p}{dt},  \label{eq:energy_diss_rate}
\end{equation}
\citep{MignardII}. If this energy was instantly converted into heat flowing out of Pluto's surface, the tidal heat flow, $F=(dE_p/dt)/(4 \pi R_p^2)$.

We can construct a basic understanding of the response of Pluto's interior to the tidal heat by comparing the heat flux from tidal dissipation to the heat flux from conduction and solid-state convection.  Solid-state convection will occur in Pluto's interior (or in its outer ice I shell, in the case of a differentiated body) if the Rayleigh number of the convecting layer,
\begin{equation}
Ra=\frac{\rho g \alpha_T \Delta T D^3}{\kappa \eta(T_i)},
\end{equation}
exceeds a critical value.  Here, $T_i$ is the temperature in the warm convecting sublayer of the ice shell, $D$ is the ice shell thickness, $\Delta T=T_i - T_s$ where $T_s \sim 40$ K is the surface temperature on Pluto, $g=0.65$ m/s$^2$ is the surface gravity, $\alpha_T = 1.56 \times 10^{-4} (T_i/250 \textrm{ K})$ K$^{-1}$, $\kappa=1.47 \times 10^{-6} (250\textrm{ K}/T_i)^2$ m s$^{-2}$ is the thermal diffusivity \citep{KirkStevenson}.  The critical Rayleigh number for convection in a Newtonian fluid with strongly temperature-dependent viscosity, $Ra_{cr} = 20.9 \theta^4$ \citep{Solomatov95}, where $\theta \approx Q^* \Delta T/(R_G T_i^2)$, and $T_i$ is the temperature in the warm convecting sub-layer of the ice shell.  The value of $T_i$ is calculated from the effective interior viscosity of the ice shell, $\eta_i$ by solving equation (\ref{eq:rheology}) for $T$.  For a uniform density pluto, $D \sim 300$ km, a rough estimate of the depth at which the lithostatic pressure exceeds 209 MPa, the location of the ice I/III phase boundary \citep{Hobbs1974}.  For a differentiated Pluto, $D=230$ km if there is no ocean, and $D=100$ km if there is an ocean (see Table \ref{table:lovenumber_params}).  

If convection occurs, the heat flow, $F_{conv}=\frac{k \Delta T}{D} Nu$, where $Nu$ is the Nusselt number, which describes the efficiency of convective heat transport relative to conductive transport alone.  For a Newtonian fluid with a strongly temperature dependent viscosity, $Nu \approx 0.53 \theta^{-4/3} Ra^{1/3}$ \citep{SM2000}.  Evaluating $Ra$, $\theta$, and $Nu$ gives an estimate for the convective heat flux,
\begin{equation}
F_{conv} = 0.53 \bigg(\frac{Q^*}{R_G T_i^2}\bigg)^{-4/3} \bigg(\frac{\rho g \alpha_T k^3}{\kappa \eta(T_i)}\bigg)^{1/3},
\end{equation}
which notably, does not depend on the thickness of the convecting layer ($D$).  

If the viscosity of the ice shell is too high to permit convection, we compare the tidal heat flux to the heat flux that could be carried by conduction alone,
\begin{equation}
F_{cond}=\frac{k \Delta T}{D},
\end{equation}
where we assume a constant thermal conductivity for the shell, $k=651/T_{ave}$ W m$^{-1}$ K$^{-1}$ \citep{PW}, and $T_{ave}=T_s+\Delta T/2$.  The net heat flux across the ice shell, $F_{net} = F - F_{cond,conv}$ is used to give a rough idea about whether the ice shell would melt or thicken during tidal evolution; $F_{net} > 0$ implies melting, and $F_{net}<0$ implies thickening.  Values of viscosity where $F_{net}=0$ are locations where the ice shell is in thermodynamic equilibrium with the tidal dissipation rate averaged over the entire orbital evolution.  Because the tidal heat flux will vary over the course of the orbital evolution, these represent only speculative stable states.

\subsection{Surface Stress Due to Orbital Evolution} \label{sec:Evolstress}
Orbital evolution in the Pluto/Charon system drives Charon's orbital radius from its starting point to a distance of 16.4 $r_p$ while changing the spin rate of Pluto. These changes in turn change the equilibrium figures of Pluto and Charon, which in turn induce stress on their surfaces. Whether the induced stress causes the formation of tectonic features on the surfaces of these two bodies depends on the magnitude of the figure change and the rate of stress buildup versus viscous relaxation.

Over the time period of orbital evolution, Pluto's spin rate changes from $\omega_{p,o}$ to $\omega_{p,f}$. The change in equilibrium spin flattening of Pluto changes by
\begin{equation}
\Delta f =\frac{h_2 r_p^3}{2 G m_p}(\omega_{p,o}^2 - \omega_{p,f}^2)  \label{eq:spinflat}
\end{equation}
where $h_2$ is a Love number, which we use at its maximum value of 5/2 to obtain a maximum stress estimate. Somewhat lower values of $h_2$ are likely due to more rigid behavior of Pluto or Charon, and would modify the stress calculated below by a factor of $2 h_2 / 5$. If the rigidity of Pluto's lithosphere at the hemispherical scale over the time period of orbital evolution is negligible, it will behave as a thin elastic shell. \citet{Melosh77} described stresses in a thin elastic shell undergoing despinning, and \citet{MatsuyamaNimmo08} derived equivalent stress equations (with the sign of the stress reversed and using a different trigonometric identity). Here we describe stresses using the positive compression convention. Based on the work of \citet{Melosh77} and \citet{MatsuyamaNimmo08}, the stresses due to despinning may be represented as
\begin{equation}
\sigma_{radial} = -\frac{\Delta f}{3}\mu\bigg(\frac{1 + \nu}{5 + \nu}\bigg)(5 - 3\cos2\lambda)  \label{eq:spinstressns}
\end{equation}
for stresses oriented north-south (radial to the spin axis) and
\begin{equation}
\sigma_{tangential} = \frac{\Delta f}{3}\mu\bigg(\frac{1 + \nu}{5 + \nu}\bigg)(1 + 9\cos2\lambda)  \label{eq:spinstressew}
\end{equation}
for stresses oriented east-west (tangential to the spin axis), where $\lambda$ is latitude, $\mu$ is the shear modulus, and $\nu$ is Poisson's ratio.

After it becomes tidally locked, Charon no longer experiences pure despinning, but instead its tidal and rotational equilibrium distortions simultaneously change as a function of distance from Pluto. This stress scenario was described by \citet{Melosh80tide}, \citet{HelfParm83}, and more recently by \citet{MatsuyamaNimmo08}.  The underlying equations are the same as (\ref{eq:spinstressns}) and (\ref{eq:spinstressew}), but the distortion of the body now has two components that can be linearly superposed: the oblate flattening around the spin axis, and the prolate flattening around the tidal axis.  Because the magnitude of the tidal and rotational distortions necessarily change together with orbital distance in a tidally locked body, the change in tidal flattening is always -3 times the change in spin flattening (cf. \citealt{Melosh80tide}; \citealt{MatsuyamaNimmo08}).  Therefore, the stress at any point can be calculated by first determining the despinning stress using (\ref{eq:spinstressns}) and (\ref{eq:spinstressew}), and then calculating the same stress, multiplied by -3, oriented radial and tangential to the tidal axis instead of the spin pole.



For both Pluto and Charon in each orbital evolution scenario, the maximum compressional stress $\sigma_c$ and tensile stress $\sigma_t$ anywhere on the surface is calculated. This is a maximum stress estimate, as it assumes that each body changed shape to match its equilibrium figure (negligible rigidity), and it assumes that all of the stress due to orbital evolution was able to accumulate without relaxing away. Like most geological materials, ice is weakest in tension, and the tensile strength of ice near Pluto's surface temperature of 40 K should be approximately 2.5 MPa \citep{Litwin2013}.  This is significantly higher than the yield strength for ice failing along pre-existing fractures \citep{Beeman} and so our choice represents a conservative value and failure in a pre-fractured lithosphere could take place at stresses an order of magnitude lower. If the maximum tensile stress on Pluto or Charon in a given orbital evolution scenario does not reach this value, we consider this case to represent ``unlikely tectonic activity." If the maximum tensile stress is above this value, tectonic activity may be possible.

To refine this assessment, we examine the rate of stress buildup over time. We calculate the Maxwell viscoelastic relaxation of stress over the entire timescale of orbital evolution $\tau$ to determine the residual stress
\begin{equation}
\sigma_{ve} = \sigma_{c,t} e^{-\frac{E\tau}{2\mu_{lid}}}  \label{eq:viscrelax}
\end{equation}
where $E$ is Young's modulus and $\mu_{lid}$ is the viscosity of the cold near-surface ice. If the residual stress is still above 2.5 MPa after viscous relaxation over the orbital evolution time period, we consider the case to represent ``likely tectonic activity."

\section{Results}
\subsection{Differentiated, No Ocean} \label{sec:no_ocean}
\begin{figure}
\centerline{\includegraphics[type=eps,ext=.epsi,read=.epsi,scale=0.5]{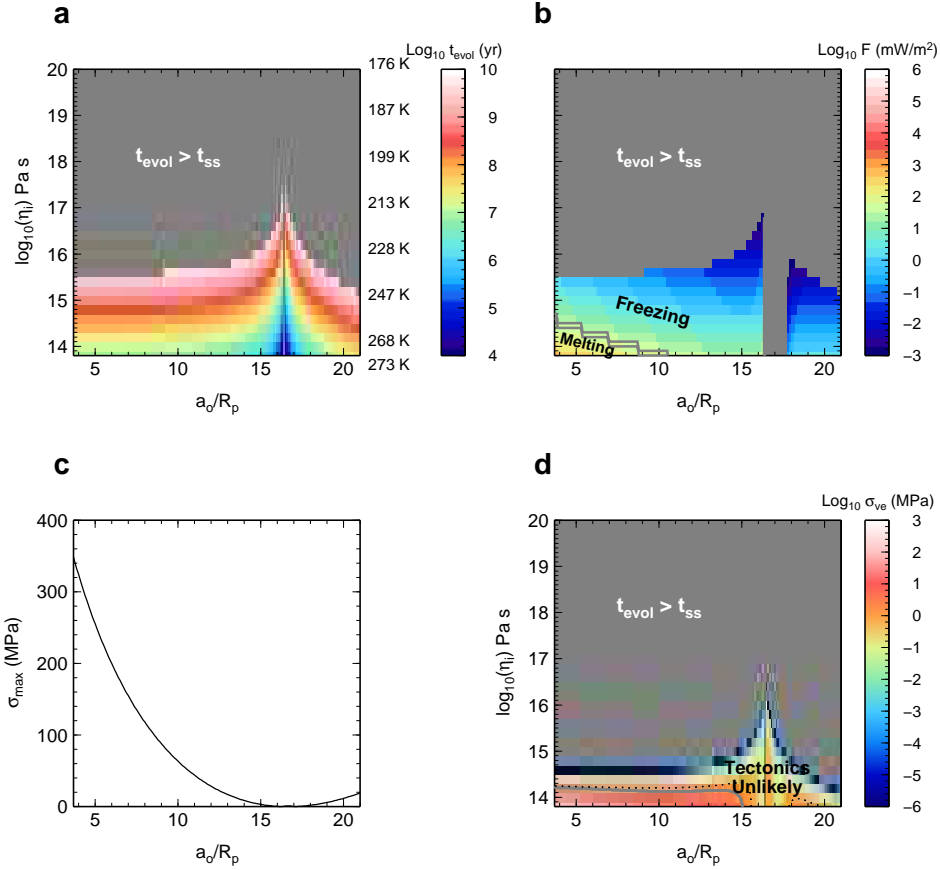}}
\caption[no_ocean]{Orbital timescales and stresses, assuming Pluto does not have an ocean (see Figure \ref{fig:interiors}b). (a) Time scale for Charon to evolve to the dual synchronous state ($a_o/R_p \sim 16.4$) depends strongly on the viscosity ($\eta_i$) of Pluto's ice shell (i.e., temperature, with approximate values shown on the right-hand axis, assuming a grain size $d=0.1$ mm).  To achieve $t_{evol}$ (colors) less than the age of the solar system, Pluto's ice shell must be warmed to $T>258$ K, assuming a minimum $d \sim 0.3$ mm. (b) Heat flux from tidal dissipation in Pluto during Charon's orbital evolution. If $F>F_{conv,max}$ (red/orange values), the ice shell will thin and become conductive. (c) Maximum tensile stress achievable due to despinning on Pluto, without regard to viscous relaxation. Larger stresses are associated with larger changes in Charon's orbital distance. (d) Maximum residual viscoelastic tensile stress remaining after time period of orbital evolution. The line between ``Tectonics unlikely'' and ``Tectonics likely'' is drawn along the 2.5 MPa contour. Dotted lines indicate a yield stress of 0.1 MPa, more appropriate for fractured ice \citep{Beeman}.  \label{fig:no_ocean} } 
\end{figure}

Figure \ref{fig:no_ocean}a illustrates the time scale for Charon to evolve from the synchronous to dual synchronous state, for the case where Pluto does not initially have an ocean (interior state depicted in Figure \ref{fig:interiors}b). The time scale for evolution is a strong function of the viscosity of the ice shell and ranges from $\sim 10^4$ years to much longer than $\sim 4.5$ Gyr, the age of the solar system. (Orbital evolution timescales longer than the age of the solar system are obviously not tenable, and so are represented as a gray shade on Figure \ref{fig:no_ocean} and subsequent figures). The evolution times for this interior model are extremely long because Pluto's ice shell is rigidly coupled to its rock core, preventing significant tidal deformation and slowing the rate at which angular momentum is transferred between the two bodies (see equation \ref{eq:dndt_dobro}).  We find that if Charon starts far from the dual-synchronous point ($a_o/R_p < 15$ or $a_o/R_p > 19$), Pluto must be warmed close to its melting point in order for Charon's orbit to evolve in $t_{evol} < 4.5$ Gyr.  
Pluto's ice shell must have an interior viscosity less than $2 \times 10^{15}$ Pa s, equal to the viscosity of ice at $T=239$ K for $d=0.1$ mm, the lowest grain size that could be expected in the ice shell \citep{gsevol}.  Assuming a more realistic $d=0.3$ mm, this viscosity is achieved only at $T=258$ K, very close to the melting point of ice.  Such a low viscosity allows enough deformation in the ice shell to permit Charon's orbit to evolve, albeit slowly.  If Charon starts close to the synchronous point ($15 \lesssim a_o/R_p \lesssim 19$), evolution times shorter than the age of the solar system are possible for viscosities up to $1.5 \times 10^{17}$ Pa s, corresponding to interior temperatures $T \sim 210$ K for the minimum grain size of 0.1 mm, or $T\sim 225$ K for $d=0.3$ mm.  

Figure \ref{fig:no_ocean}b illustrates the heat flux from tidal dissipation, averaged over Charon's orbital evolution.   We find that the heat flows range from $F=10^{-3}$ to $\sim 100$ mW m$^{-2}$, although the vast majority of orbital scenarios give heat flows of order tens of milliwatts per meter squared.  For the vast majority of orbital evolution scenarios and interior viscosities, the amount of tidal heat deposited in Pluto's ice shell will be small enough to allow Pluto to remain solid (and avoid forming an ocean) during Charon's orbital evolution.

Figures \ref{fig:no_ocean}c and \ref{fig:no_ocean}d illustrate the maximum stresses on the surface of Pluto during the orbital evolution.  The maximum stress achieved is large $\sigma_{max} =340$ MPa, but decreases as a function of $a_o/R_P$. Tectonic activity is ``likely'' on Pluto (as defined at the end of section \ref{sec:Evolstress}) only if the ice shell has a very low viscosity $\eta_i < 2 \times 10^{14}$ Pa s and $a_o/R_p < 15.2$ or $a_o/R_p > 18.2$.

Thus, we consider it likely that Charon can evolve into the dual synchronous state even if Pluto does not have an ocean and its outer ice shell is mechanically coupled to its much stiffer rock core.  However, Pluto will only display evidence of ancient tectonic resurfacing if its interior viscosity is quite low, corresponding to a warm ice shell.  For $a_o/R_p \sim 3.4$ to 10, model plutos with very warm ice shells could likely experience runaway melting from tides, leading to the formation of an ocean.  For $a_o \sim 10$ to 15.2, a warm ice shell on Pluto can be tectonically active without melting from tidal dissipation.  

\subsection{Undifferentiated} \label{sec:undiff}
\begin{figure}
\includegraphics[type=eps,ext=.epsi,read=.epsi,scale=0.65]{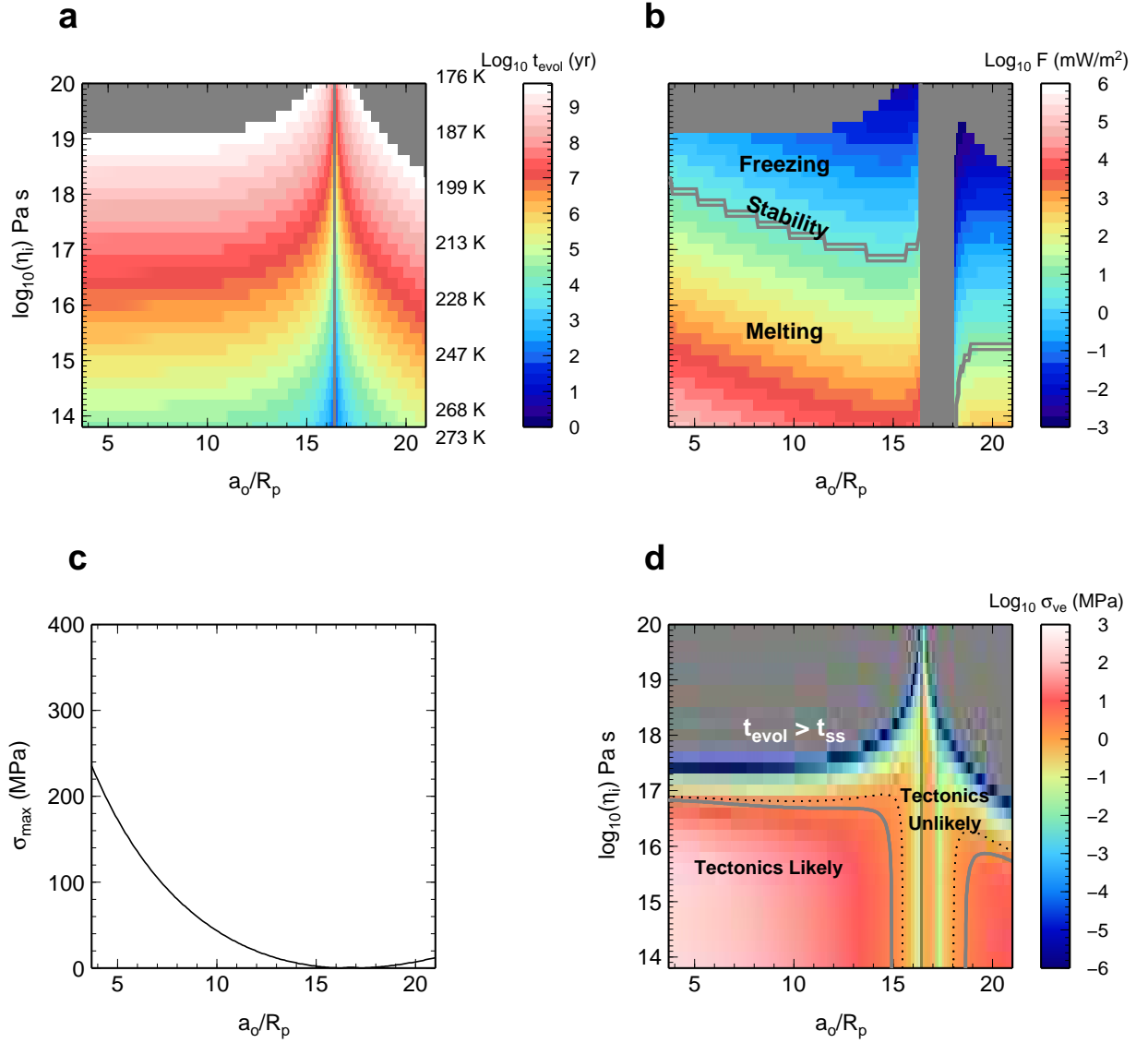}

\caption[undiff]{Similar to Figure \ref{fig:no_ocean}, but for a uniform-density (undifferentiated) Pluto (interior structure shown in Figure \ref{fig:interiors}c). (a) Time scale for orbital evolution. (b) Heat flux from tidal dissipation in Pluto. (c) Maximum despinning tensile stress achievable on Pluto. (d) Residual viscoelastic tensile stress after orbital evolution. \label{fig:undiff}}
\end{figure}

Figure \ref{fig:undiff} illustrates our results for an undifferentiated Pluto.  Here, the entire interior of Pluto has a rheology appropriate for water ice, which allows it to experience more tidal deformation and dissipation than a differentiated model pluto without an ocean whose ice shell is rigidly coupled to its rocky core.  We find that an undifferentiated Pluto/Charon system can reach the dual synchronous state in a time scale less than the age of the solar system if its interior viscosity is less than $10^{19}$ Pa s, or $T=173$ K for $d=0.1$ mm, or $T=198$ K for $d=0.3$ mm (see Figure \ref{fig:undiff}a).  

An undifferentiated Pluto experiences significantly more tidal dissipation than its differentiated counterpart, even if it never has an ocean.  Figure \ref{fig:undiff}b shows the estimated tidal heat fluxes, and the range of interior viscosities that permit freezing or melting of the ice shell.  Initially high-viscosity (cold) shells will remain cold and could thicken.  Ice shells that are initially warm, with $\eta_i < 10^{16}$ to $10^{17}$ Pa s, or $T \sim 200$ to 230 K, depending on grain size, will experience strong tidal dissipation in excess of the amount that could be removed by conduction and/or convection, which could lead to melting.  Because runaway differentiation may be possible in Pluto (see Section \ref{sec:interiors}), we consider that the onset of melting would trigger complete differentiation of the body in addition to thickening of the outer ice shell and formation of an ocean; the interior state of Pluto evolves to that depicted in Figure \ref{fig:interiors}a.  Once the ice shell began to melt, the amount of tidal dissipation would decrease since the thickness of the shell had decreased \citep{CRPTidal, CRPTidalCorrection}.  Eventually an equilibrium between tidal heat and conductive/convective heat transport may be achieved (e.g., \citealt{Moore2006}).  

The maximum stress experienced is $\sigma_{max} \sim 250$ MPa, far in excess of the strength of water ice. Tectonic activity is ``likely'' on Pluto if its ice shell has a viscosity $\eta_i < 7 \times 10^{16}$ Pa s and $a_o/R_p < 15$ or $a_o/R_p > 18.6$. Because this viscosity is associated with tidal dissipation that would lead to melting, an undifferentiated Pluto is incompatible with tidally-driven tectonic activity.

\subsection{Differentiated, With Ocean} \label{sec:diff_ocean}
\begin{figure}
\includegraphics[type=eps,ext=.epsi,read=.epsi,scale=0.65]{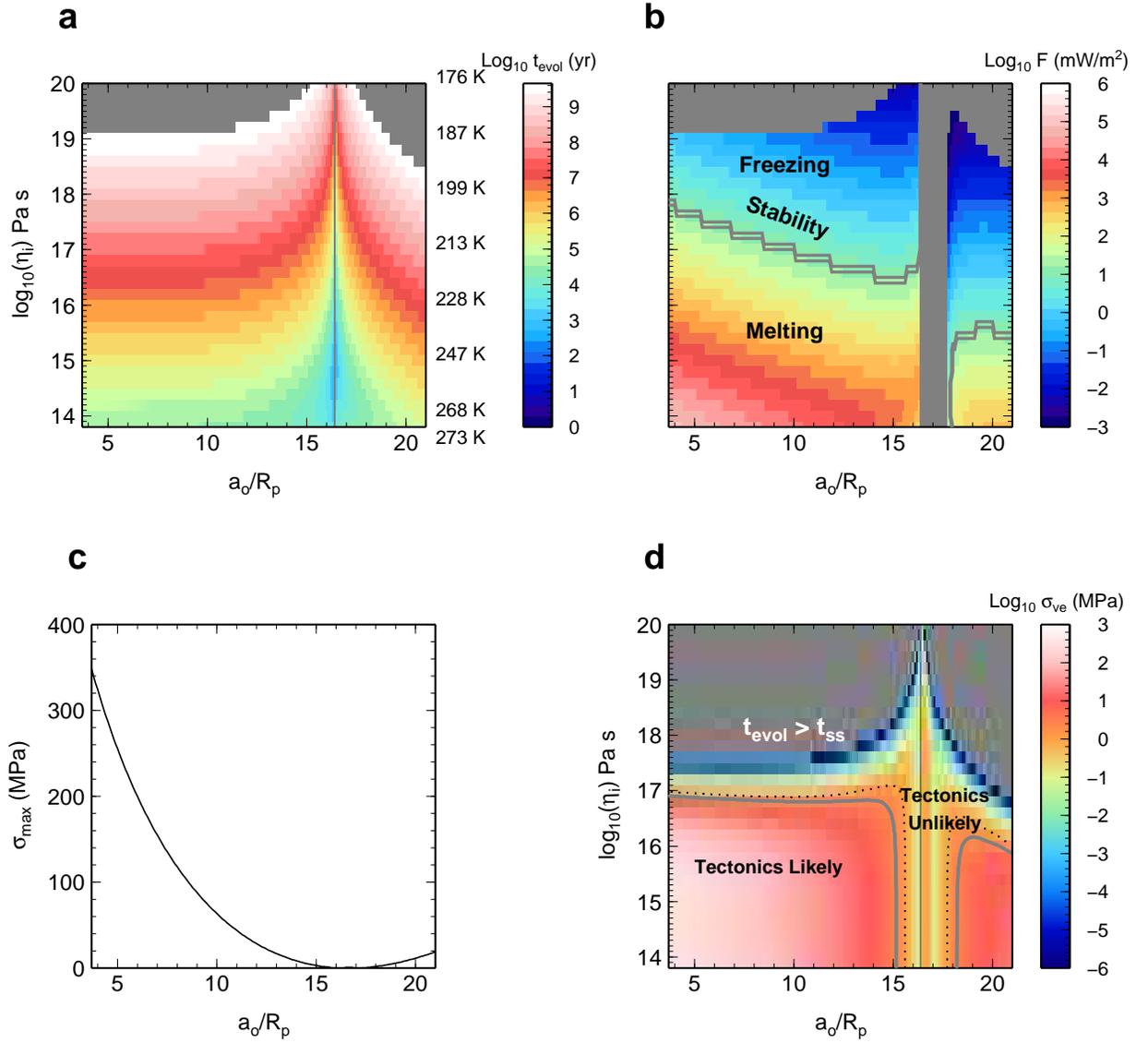}

\caption[undiff]{(a) Similar to Figure \ref{fig:no_ocean}, but for a differentiated Pluto with an ocean (interior structure shown in Figure \ref{fig:interiors}a). (a) Time scale for orbital evolution. (b) Heat flux from tidal dissipation in Pluto. (c) Maximum despinning tensile stress achievable on Pluto. (d) Residual viscoelastic tensile stress after orbital evolution. \label{fig:ocean}}

\end{figure}

Figure \ref{fig:ocean} illustrates the evolution time scale, heat flux, and stresses in Pluto during Charon's orbital evolution, assuming that Pluto is fully differentiated and has a liquid water ocean.  Similar to the undifferentiated case, the orbital evolution time scale is less than the age of the solar system if $\eta_{i}< 10^{19}$ Pa s.  Tidal heat can be removed by conduction and/or convection if $\eta_i \sim 10^{16}$ to $10^{18}$ Pa s, roughly corresponding to $T \sim 200$ to 245 K.  The presence of a subsurface ocean permits stresses $\sigma_{max} \sim 350$ MPa. Tectonic activity is ``likely'' on Pluto if its ice shell has a viscosity $\eta_i < 8 \times 10^{16}$ Pa s and $a_o/R_p < 15.2$ or $a_o/R_p > 18.2$.

\subsection{Stress on Charon}
Stress due to synchronized lowering of tidal and rotational bulges (tidal recession) was calculated on Charon for each Pluto scenario discussed above.  Theoretical maximum tensile stress on Charon due to tidal recession is $\sigma_{max} \sim 250$ MPa, though the thin elastic shell assumption may be less valid for Charon than for Pluto, as Charon is not strongly heated. Tectonic activity is ``likely'' on Charon for $a_o/R_p < 13.4$ and for the same ranges of Pluto's interior viscosity that made activity likely on Pluto for any of the three interior models. This is because it is Pluto's interior that controls the rate of orbital evolution, and thus the rate of stress accumulation versus viscous relaxation.
 
\section{Discussion}\label{sec:conclusions}
The \emph{New Horizons} mission has sparked new interest about the thermal evolution of Pluto and its moon Charon, in particular, the possibility of tectonic resurfacing on both bodies and the likelihood that these bodies may have had liquid water oceans, either now or in the past.  In this work, we have used simple tidal models to calculate the magnitude of stresses expected on the surfaces of Pluto and Charon during their evolution into the dual-synchronous state.  During this time, Pluto de-spins as Charon recedes from Pluto, potentially leading to enormous stresses on the lithospheres of both bodies.  

We conclude that if \emph{New Horizons} uncovers evidence for ancient tectonic activity consistent with despinning on Pluto and/or tidal recession on Charon, the most self-consistent explanation is that Pluto had an ocean during the time period of Charon's orbital evolution.  Despinning stresses on Pluto would manifest as normal faults at the poles, strike-slip features in the mid-latitudes, and equatorial thrust faults \citep{Melosh77}.  Stresses on Charon arising from the collapse of its tidal bulge would create normal faults at the poles, strike-slip faults in the mid-latitudes, with a zone of thrust faults at the sub-Pluto point \citep{Melosh80tide}.  Concurrent volume changes in the interiors of Pluto or Charon could add to these predicted stress states and significantly shift the boundaries of the tectonic regions toward or away from the poles or tidal axis. If \emph{no} tectonic features are observed, this could indicate an extremely cold Pluto pre- or post-impact (and thus a very long timescale for orbital evolution), or that tectonic features had been removed or buried by other processes.  Tectonic features formed by the post-impact evolution could have been obscured by, e.g., later flooding by cryomagmas.  It is also possible that infall of Pluto's tenuous atmosphere may subdue landforms created by ancient tectonics.

We find that the orbital evolution time scale for Charon depends sensitively on the interior thermal state of Pluto.  This is because the tidal quality factor of Pluto, which controls the rate at which angular momentum is transferred from Pluto's spin to Charon's orbit, depends on $Q$, which in turn depends on the viscosity of its interior.  A cold, highly viscous (``stiff'') Pluto will shed angular momentum more slowly than a warm, low-viscosity Pluto.  If Pluto is differentiated and has a liquid water ocean, its ice shell is no longer rigidly coupled to its rock core, permitting large tidal deformations and rapid tidal evolution even if the ice shell has a high viscosity.  
For Charon to evolve to the dual synchronous state on a time scale less than the age of the solar system, a fully differentiated Pluto with an ocean (Figure \ref{fig:ocean}) must be warmed post-impact to a temperature $T \gtrsim 170$ to 190 K, so that the viscosity of the convecting portion of its ice shell, $\eta_i < 10^{19}$ Pa s.  Similar conditions must be achieved in an undifferentiated Pluto.  However, we consider it likely that Pluto is differentiated either before or during Charon's tidal evolution because the energy liberated during accretion is sufficient to melt its ice, Pluto is unstable to runaway differentiation, and its high mean density implies a high rock fraction in the interior and thus, ample radiogenic heat.  
If Pluto did not have an ocean during Charon's orbital evolution, its interior must be warmed to $\sim 240$ to 260 K for Pluto and Charon to evolve to the dual synchronous state.  Temperatures this high would lead to melting, particularly if Pluto has even just a small amount of ammonia or low-eutectic salt mixed in with its ices.  

While it is possible for orbital evolution to drive tectonics on Pluto and Charon for any of the assumed interior models for Pluto,  the thermal/orbital pathway of the system depends sensitively on the interior states of Pluto and Charon, which may change over time.  For example, in the differentiated model without an ocean, tectonics can only occur if Charon starts close to its current location, and Pluto's ice is very close to the melting point (but not molten).  In the undifferentiated model, tectonic activity occurs if the ice has a low viscosity, but this same low viscosity ice will dissipate enough heat to begin melting and possibly trigger runaway differentiation.  In the differentiated Pluto model with an interior ocean, tectonic activity occurs above an interior that has the right viscosity for a floating, conducting or convecting ice shell.  Thus, more detailed calculations (e.g., \citealt{RobuchonNimmo}) will be warranted once the \emph{New Horizons} mission has returned its data.  


\section*{Acknowledgements}
Author Barr acknowledges support from OPR NNX12AL22G.  We thank R. M. Canup, R. T. Pappalardo, and S. A. Stern for helpful discussions.


\begin{table}
\begin{tabular}{l  ll}
\hline
Parameter &  Pluto & Charon  \\
\hline
Mass$^a$ & $1.304 \times 10^{22}$ kg & $1.520 \times 10^{21}$ kg \\
Radius$^a$ & 1147 km & 606 km  \\
Mean Density$^a$ & 2060 kg m$^{-3}$ & 1630 kg m$^{-3}$ \\
Surface Gravity & 0.66 m/s$^2$ & 0.28 m/s$^{2}$ \\
Present Orbital Period$^a$ & --   & 6.38270 days\\
Present Spin Period$^a$ & $1.139\times 10^{-5}$ s$^{-1}$ &  $1.139\times 10^{-5}$ s$^{-1}$\\
Ice Density &  1000 kg m$^{-3}$ & 1000 kg m$^{-3}$ \\ 
Rock Density &  3000 kg m$^{-3}$ & 3000 kg m$^{-3}$ \\ 
Rock Mass Fraction &  0.81 & 0.58 \\
Rock Volume Fraction & 0.53 & 0.315 \\
\hline
Moment of Inertia Factor &  & \\
Differentiated  & 0.328 & 0.317 \\
Undifferentiated & 0.4 & 0.4 \\
\hline
\end{tabular}
\label{table:satellite_params}
\caption{Properties of the Pluto/Charon System.  $^a$ From four-body orbital solution of \citet{Tholen2008}}
\end{table}

\begin{table}
\caption{Physical and Rheological Properties of a Differentiated Pluto}
\begin{tabular}{l l   ll}
\hline
Layer & Parameter &  Pluto, No Ocean & Pluto With Ocean  \\
\hline
Rocky Core & Radius &  917 km & 917 km \\
			  & Density  & 3000 kg m$^{-3}$& 3000 kg m$^{-3}$ \\
			  & Young's Modulus  & $2.5 \times 10^{11}$ Pa & $2.5 \times 10^{11}$ Pa\\
			  & Poisson's Ratio & 0.25 & 0.25 \\
\hline
Ocean		  & Thickness  & -- & 130 km \\
			  & Density	 & --& 1000 kg m$^{-3}$ \\
			  & P-Wave Velocity  & --& 1.45 km s$^{-1}$ \\
\hline		  
Ice Shell	  & Total Thickness &230 km & 100 km \\
			  & Rigid Lid Thickness & -- & 15 km \\
			  & Lid Viscosity & $10^{24}$ Pa s & $10^{24}$ Pa s\\
			  & Sublayer Thickness  & 219 km & 85 km \\
		  	  & Sublayer Viscosity & variable & variable \\
\hline
\end{tabular}
\label{table:lovenumber_params}
\end{table}

\clearpage

\begin{figure} 
\centerline{\includegraphics[type=eps,ext=.epsi,read=.epsi,scale=.3]{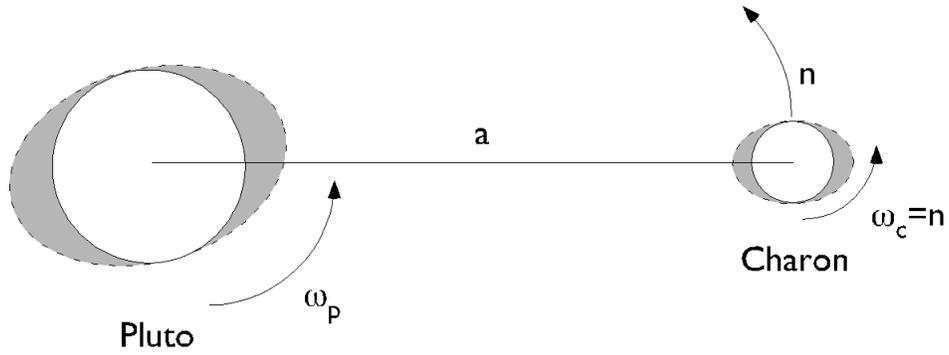}}
\caption[Pluto.Charon.Orbit.tif]{Schematic of the orbital state of Pluto and Charon during
the system's evolution assumed in our study.  After the P/C-forming impact, Charon is launched 
into an eccentric orbit around Pluto, but rapidly evolves into a synchronous orbit where its spin frequency
$(\omega_c)$ is equal to its orbital frequency $(n)$.  Pluto spins rapidly with frequency $\omega_p$.  The
system continues to evolve over a long time scale, $\tau_{evol}$, as the spin angular momentum of Pluto
is transferred to Charon, changing its orbital semi-major axis, $a$.  Tides are raised and lowered in Pluto
with a frequency $\omega_{tidal} = \omega_p-n$.  In the present-day dual synchronous state,
$\omega_p = \omega_c = n = 1.13 \times 10^{-5}$ s$^{-1}$, and $a=16.4 R_p$.
\label{fig:Pluto.Charon.Orbit.epsi}}
\end{figure}


\clearpage


\clearpage


\clearpage


\clearpage

\end{document}